\documentclass[aps,amssymb,amsmath,twocolumn,prx]{revtex4}
\usepackage{graphicx}
\usepackage{amssymb}
\usepackage{amsmath}
\usepackage{bm}
\usepackage{xcolor} 
\usepackage{array}
\usepackage{multirow}
\usepackage{tabularx}
\usepackage{booktabs}
\usepackage{textcomp}

\usepackage{bbm}
\usepackage{cancel}
\usepackage{comment}

\newcommand{\ket}[1]{\left|#1\right\rangle}
\newcommand{\abs}[1]{\left|#1\right|}

\begin{document}
\title{Quantum annealing of Cayley-tree Ising spins}

\author{Yunheung Song}
\altaffiliation{Present address: Department of Physics, University of Wisconsin-Madison, 1150 University Avenue, Madison, Wisconsin 53706, USA}
\author{Minhyuk Kim}
\author{Hansub Hwang}
\author{Woojun Lee}
\altaffiliation{Present address: Department of Computer Science and Engineering, Seoul National University, Seoul 08826, Republic of Korea }
\author{Jaewook Ahn}
\thanks{jwahn@kaist.ac.kr}
\affiliation{Department of Physics, Korea Advanced Institution of Science and Technology, Daejeon 34141, Republic of Korea}
\date{\today}

\begin{abstract} \noindent
Significant efforts are being directed towards developing a quantum annealer capable of solving combinatorial optimization problems. The challenges are Hamiltonian programming in terms of high dimensional qubit connectivity and large-scale implementations. Here we report quantum annealing demonstration of Ising Hamiltonians programmed with up to $N=22$ spins mapped on various Cayley-tree graphs. Experiments are performed with a Rydberg-atom quantum simulator, in which rubidium single atoms are arranged in three dimensional space in such a way that their Rydberg atoms and blockaded strong couplings respectively represent the vertices and edges of each graph. Three different Cayley-tree graphs of $Z=3$ neighbors and of up to $S=4$ shells are constructed, and their ground-state phases and N\'{e}el's order formations are probed. The anti-ferromagnetic phase in regular Cayley trees and frustrated competing ground-states in a dual-center Cayley tree are directly observed,  demonstrating the possibilities of high-dimensional qubit connections in quantum simulators.
\end{abstract}

\maketitle

\section{Introduction} 
\noindent
In recent years, quantum annealers have received significant attention because of their potentials in solving complex computational problems which are often intractable with non-quantum computational methods~\cite{Farhi2001,Das2008,Albash2018,Hauke2020}. Quantum annealing is a procedure of making Hamiltonian $H(t)$ of a quantum many-body system adiabatically evolve from $\hat H_i$ to $\hat H_f$, 
\begin{equation}
\hat{H}(t)=\hat{H}_i -\frac{t}{t_f}(\hat{H}_i-\hat{H}_f), \label{H1}
\end{equation}
so that the quantum state $\ket{\Psi(t)}$ initially prepared in the ground state of the former reaches the ground state of the later. Quantum annealers are considered with superconducting qubits~\cite{Johnson2011,Bunyk2014,Rosenberg2017,HarriganNP2021} and trapped-ion qubits~\cite{Korenblit2012,Islam2013,Richerme2013,Hauke2015}, aiming for various combinatorial optimization problems such as quantum simulations~\cite{Babbush2014}, classfications~\cite{Neven2008}, planning~\cite{Rieffel2015}, etc.  While many efforts in quantum annealing are being focused on large-scale implementations~\cite{Young2013,Boixo2014,Pudenz2014,Ebadi2020,Scholl2020} towards quantum speedup~\cite{Altshuler2010,Somma2012,Ronnow2014,Muthukrishnan2019}, here we explore the possibility of high-dimensional qubit connectivities. There are theoretical proposals emphasizing and thus utilizing qubit connectivities for NP-hard optimization problems: for examples, Lechner-Hauke-Zoller scheme~\cite{LHZ2015,Glaetzle2017} proposed a quantum annealing architecture for all-to-all connectivities using local interactions; and quantum optimization protocols are considered, e.g., for maximum independent set problems utilizing the nature of long-range couplings especially in Rydberg-atom quantum simulators~\cite{Pichler2018}. In the context relevant to the present work, Rydberg-atom quantum simulators~\cite{Weimer2010,Browaeys2020} draw attention, because of their high tunability in qubit connectivities~\cite{Lee2016,Kim2016,Barredo2016,Endres2016,Barredo2018} as well as many-body controllability in adiabatic processes~\cite{Pohl2010,Schauss2015,Bernien2017,Lienhard2018,Beterov2020}.
 
\begin{figure*}[t]
\centerline{\includegraphics[width=1.00\textwidth]{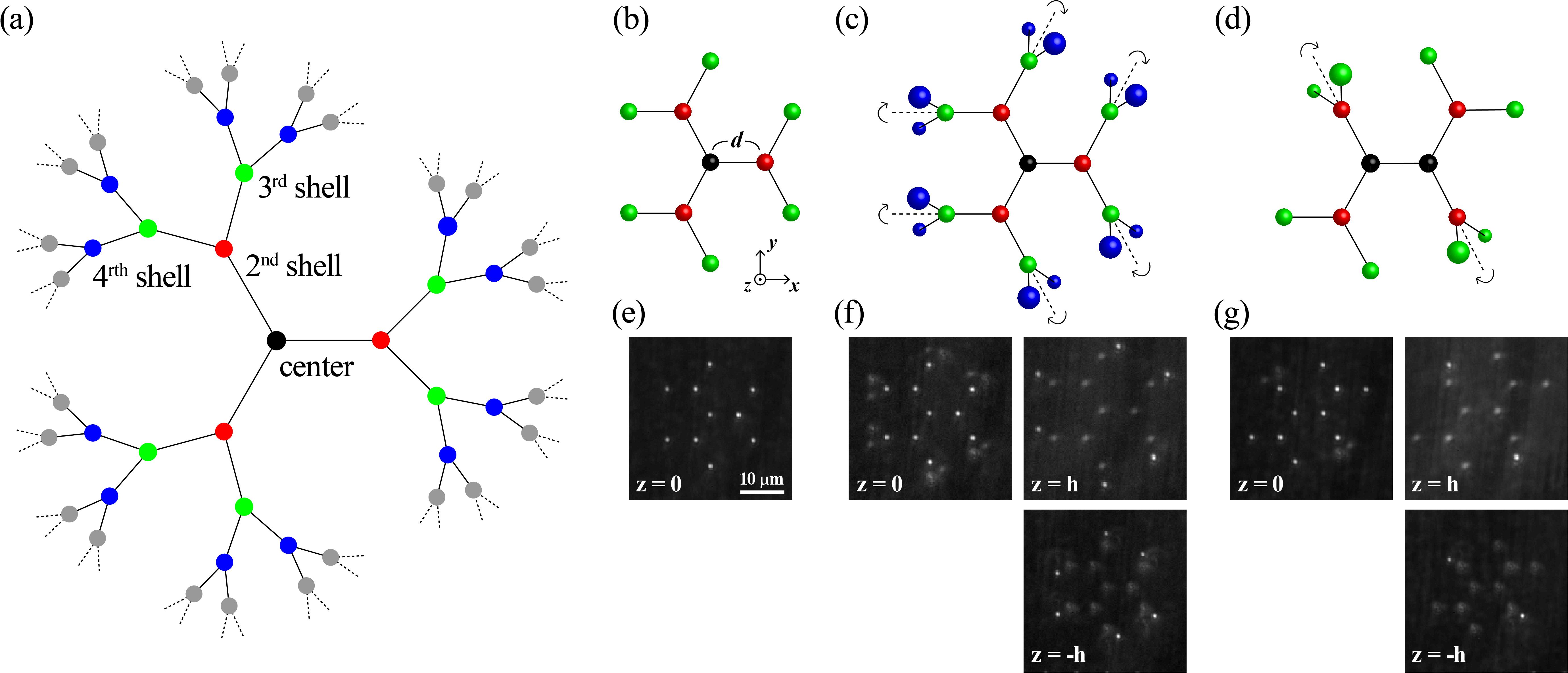}}
\caption{%{\bf Cayley tree.} 
(a) A generic Cayley-tree graph of coordination number $Z=3$, with vertices and edges representing atoms and Rydberg-blockaded atom pairs, respectively. (b) A three-shell Cayley tree ($G_{10}$) of 10 atoms of inter-atom distance $d$. (c) A four-shell Cayley tree ($G_{22}$) constructed in three planes at $z=0, \pm h$ ($h=d/1.2$), where the last-shell branches are rotated by $72^\circ$ to avoid unwanted couplings. (d) A dual-center Cayley tree ($G_{14}$) of 14 atoms. (e,f,g) Plane-by-plane fluorescence images of the corresponding atom arrangements.}
\label{fig1}
\end{figure*}

In this work, we consider, as a prototypical fractal structures, Cayley tree graphs of neutral atom arrangements, in which atoms and strongly interacting atom pairs respectively represent vertices and edges of the graphs (see Fig.~\ref{fig1}). Cayley trees are homogeneous and isotropic tree graphs of a fixed number of edges and no loop~\cite{Bethe1935,Ostilli2012}. Their infinite version is Bethe lattice widely used in various physics areas as a fundamental theoretical platform, often providing exactly solvable models in classical and quantum problems~\cite{Baxter1982}. In experiments to be described below, we construct atomic Cayley graphs of coordination number $Z=3$ and shell number up to $S=4$, as in Fig.~\ref{fig1}(a), and use a Rydberg-atom quantum simulator to directly probe their Ising-Hamiltonian phases.

\section{Cayley-tree atom arrangements in three-dimensional space} 
\noindent
Neutral atoms (rubidium, $^{87}$Rb) are arranged in three-dimensional (3D) space with optical tweezers (far-off resonant optical dipole traps)~\cite{Barredo2018,Kim2020} (see Sec.~\ref{technical} for technical details). Three different Cayley-tree graphs are constructed. The first one is the three-shell ($S=3$) Cayley tree, which can be denoted by $G_{10}=(0s)(1s)^3(2s)^6$, having one in the first (center) shell, three in the second shell, and six in the third shell, as in Fig.~\ref{fig1}(b). The second one is the four-shell ($S=4$) Cayley tree, $G_{22}=(0s)(1s)^3(2s)^6(3s)^{12}$, having ten atoms in $G_{10}$ and twelve in the fourth shell, as in Fig.~\ref{fig1}(c).  And the last one is $G_{14}=(0s)^2(1s)^4(2s)^8$, a dual-center Cayley tree, having two first-shell atoms, four second-shell atoms, and eight third-shell atoms, as in Fig.~\ref{fig1}(d).

In our atom arrangements, each atom represents a vertex of a Cayley-tree graph and each pair of strongly-interacting atoms an edge. Non-connected atoms are supposed to interact with each other much weakly than connected atoms. The Hamiltonian of these atoms being coherently and simultaneously excited to a Rydberg energy state is given (in unit of $\hbar=1$) by
\begin{equation}
\hat{H}= \frac{1}{2} \sum_{j=1}^N\left\{\Omega\hat\sigma^{(j)}_x - \Delta\hat\sigma^{(j)}_z\right\}+ \sum_{j<k} U_{jk} \hat n^{(j)}\hat n^{(k)},
\label{H}
\end{equation}
where $N$ is the number of atoms, $\Omega$ is the Rabi frequency, $\Delta$ is the detuning, and $U_{jk}=C_6/|\vec{r}_j-\vec{r}_k|^6$ is the  pairwise atom interaction in the van der Waals interaction regime~\cite{Bernien2017,Lienhard2018}. Pauli operators $\hat\sigma_{x,z}$ are defined for a pseudo spin $1/2$ system composed of the ground state $\ket{\downarrow}=\ket{5S_{1/2}, F=2, m_f=2}$ and Rydberg state $\ket{\uparrow}=\ket{71S_{1/2},F'=3,m_F'=3}$ of each atom, and $\hat{n}=(1+\hat\sigma_z)/2$. For the strong and weak interactions of connected and nonconnected atom pairs, respectively, we set the distances of connected atoms the same and within the Rydberg-blockade radius, i.e., $|\vec{r}_j-\vec{r}_k|=d<r_b\equiv (C_6/\hbar\Omega)^{1/6}=9.8~{\rm \mu m}$, for $(j,k)\in E$ (the edge set of a graph $G$), and the distances of all others are $|\vec{r}_j-\vec{r}_k|>r_b$.

The two-dimensional (2D) arrangement of $G_{10}$ is made with $d<r_b<\sqrt{3}d$ to satisfy the above condition, in which, as shown in Fig.~\ref{fig1}(e), the last-shell atoms of different branches are separated more than $\sqrt{3}d$, having at most 1/27 times smaller  interactions than $U\equiv C_6/d^6$ of the connected atom pairs. 
However, $G_{22}$ and $G_{14}$ cannot be planar, because in 2D arrangements, atoms of different outermost branches are too close, requiring nonplanar, three-dimensional (3D) arrangements. As shown in Fig.~\ref{fig1}(c), we rotate the last-shell branches of $G_{22}$ by an angle of $2\pi/5$ about the axes along the previous branches, so that all the last-shell atoms are well separated (more than $\sqrt{3}d$) from each other. In $G_{14}$, some last-shell branches are rotated similarly, as shown in Fig.~\ref{fig1}(d). As-constructed atom arrangements are shown in Figs.~1(e-g) respectively for $G_{10}$, $G_{22}$, and $G_{14}$. 

\section{Phase diagrams of Cayley-tree Ising spins} 
\noindent
With atoms arranged on one of the above Cayley-tree graphs, we perform quantum annealing to find the ground state of a target Hamiltonian. The atoms are initially prepared in $\ket{\downarrow\downarrow\cdots\downarrow}$ (the paramagnetic down spins in Phase I) under the Hamiltonian conditions of  $\Delta \gg U$ and $\Omega=0$ in $\hat H$. The target (final) Hamiltonian  is 
\begin{eqnarray}
\hat{H}_G (U,\Delta_f)=U \sum_{(j,k)\in E}\hat n^{(j)}\hat n^{(k)} -\frac{\Delta_f}{2} \sum_{j=1}^N \hat\sigma^{(j)}_z
\label{HG}
\end{eqnarray}
where $E$ is the edge set of $G\in \{ G_{10}, G_{22}, G_{14}\}$, $\Delta_f$ is the final detuning, and no couplings are assumed for unedged atom pairs. We note that $\hat H_G$ is an Ising spin-glass Hamiltonian, given by
\begin{eqnarray}
\hat{H}_G = J \sum_{(j,k)\in E} \hat \sigma_z^{(j)}\hat \sigma_z^{(k)} 
+   h_z^C    \sum_{j\in C}  \hat \sigma^{(j)}_z + h_z^V   \sum_{j\in V }  \hat \sigma^{(j)}_z,  
\label{HG2}
\end{eqnarray}
where $J=U/4$ is the coupling, $h_z^C= 3U/4-\Delta_f/2$ and $h_z^V=  U/4-\Delta_f/2$ are the local fields, and $C$ and $V$ denote the vetices in the core (inner shells) and valence (outermost) shell, respectively, of a Cayley-tree graph $G(E,C+V)$. In general, as the ground-states of an Ising Hamiltonioan depend on the specific atom arrangement of $G$, finding such arrangement-specific ground-states of an arbitrary Ising-spin graph is an NP-complete problem~\cite{Lucas2014}. 

\begin{figure}[tb]
\centerline{\includegraphics[width=0.50\textwidth]{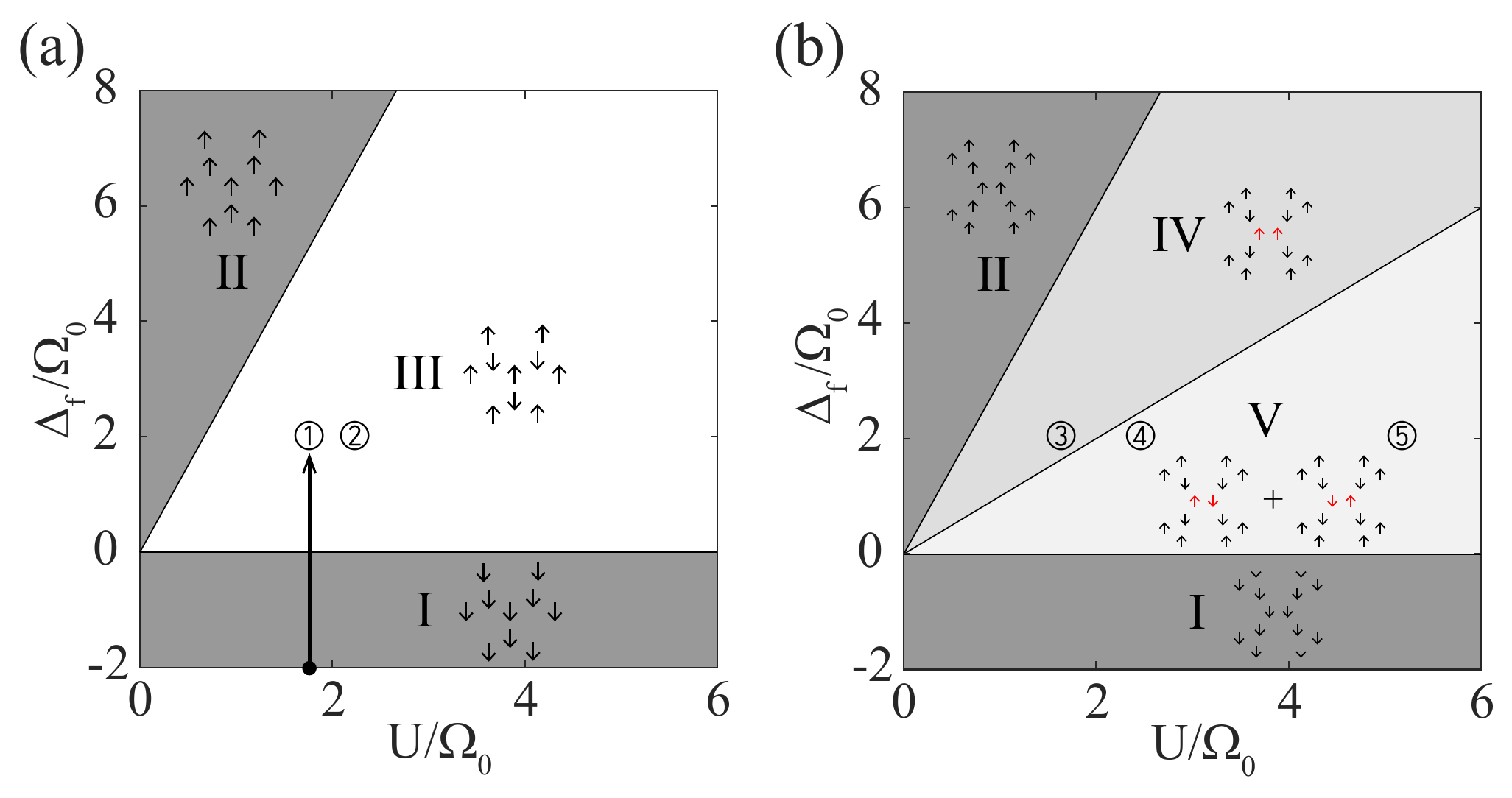}}
\caption{Phase diagrams of Cayley-tree Ising Hamiltonians (a) $\hat H_{G_{10}}$ and $\hat H_{G_{22}}$ and (b) $\hat H_{G_{14}}$, in which ground-state spin configurations are paramagnetic down spins (Phase I), paramagnetic up spins (Phase II), antiferromagnetic phase (Phase III), and antiferro-like phases with center-spins $\ket{\uparrow\uparrow}$ (Phase IV) and $(\ket{\uparrow\downarrow+\downarrow\uparrow})/\sqrt{2}$ (Phase V).
Circled numbers indicate the Hamiltonian parameters of the experimental data in Figs.~\ref{fig3} and \ref{fig4}: {\textcircled{\small 1}} ($U/\Omega_0$, $\Delta_f/\Omega_0$)=(1.82, 2) in Phase III of $G_{10}$, 
{\textcircled{\small 2}} (2.25, 2) in Phase III of $G_{22}$,  {\textcircled{\small 3}} (1.67, 2) in Phase IV of $G_{14}$, 
{\textcircled{\small 4}} (2.70,2) and {\textcircled{\small 5}} (5.41,2) in Phase V of $G_{14}$. $\Omega_0=1.1$~$(2\pi)$MHz in all experiments. 
}
\label{fig2}
\end{figure}

However, Cayley-tree graphs allow heuristic understanding of their phase diagrams as follows: Cayley trees have more atoms on the valence shell than on the rest (the inner shells) of a tree. Therefore, with Hamiltonian $\hat H_G$ in Eq.~\eqref{HG} (of positive coupling, $U>0$, in our consideration), the valence spins are all aligned either up or down depending upon $\Delta_f>0$ or $\Delta_f<0$, respectively, resulting in the $\Delta_f=0$ phase boundary. For $\Delta_f<0$, inner-shell spins, adjacent to the valence spins, also favor down spins, as their couplings, $U\hat n^{(j)} \hat n^{(k)}$, to adjacent outer-shell down spins are zero, which sequentially results in all down spins (so paramagnetic down spins, Phase I). For $\Delta_f>0$, the single-spin flipping energy ($\Delta_f$) competes with adjacent anti-ferromagnetic couplings. For $\Delta_f>3U$, the former is always higher than the latter (of max $3U$), so all spins are up (paramagnetic up spins, Phase II). In between, $0<\Delta_f<3U$, the anti-ferromagnetic coupling is stronger, thus favoring the anti-ferromagnetic phase (Phase III) of shell-by-shell alternating spins. The resulting $\hat H_G$ phase diagram is shown for $G_{10}$ in Fig.~\ref{fig2}(a), in which the anti-ferromagnetic phase (Phase III) has up spins on the first and thrid shells and down spins on the second. The phase diagram for $G_{22}$ differs from $G_{10}$ only in Phase III, where the odd-number shells are down spins and the even-number shells are up spins. In short, the phase diagram of $\hat H_G$ (of positive $U$) for regular Cayley-graphs ($G_{10}$ and $G_{22}$ in our consideration) has the para-antiferro phase boundaries determined by $\Delta_f=3U$ and $\Delta_f=0$, so the anti-ferromagnetic ordering is expected in the region $0<\Delta_f<3U$ (Phase III), as in Fig.~\ref{fig2}(a).

The phase diagram for $G_{14}$, the dual-center Cayley tree, is a little more complex than regular Cayley trees, because of the frustrations of the center spins. Exact diagonalization of $\hat H_{G_{14}}$ finds the phase diagram as shown in Fig.~3(b), in which the paramagnetic phases (Phases I and II) are the same as $G_{10}$, but Phase III is split to Phases IV and V, of respective center-spin configurations of $\ket{\uparrow\uparrow}$ and $(\ket{\uparrow\downarrow+\downarrow\uparrow})/\sqrt{2}$. The new phase boundary is $U=\Delta_f$, along which the single-spin flipping cost  equals the additional frustration cost of the center spins. It is noted that, in Phase V, other energy-degenerate ground states, with respective center-spin configurations of $\ket{\uparrow\downarrow}$, $\ket{\downarrow\uparrow}$, and $(\ket{\uparrow\downarrow-\downarrow\uparrow})/\sqrt{2}$, are all dark states for a quantum annealing performed to the given initial state, due to the symmetey of $\hat H$ in Eq.~\eqref{H}.

\section{Experimental verification of Cayley-tree Ising Phases} 
\noindent
Experiments are performed to verify the antiferro-like ground states (the Phases III, IV, and V) of Cayley-tree Ising spins (see Sec.~\ref{technical} for technical details). Quantum annealing is proceeded with three stages of time evolution, along a vertical control path from Phase I to either Phase III, IV or V in Fig.~\ref{fig2}. In the first stage ($0<t<0.1t_f$), Rabi frequency is adiabatically turned on from $\Omega_i=0$ to $\Omega_0=1.1 (2\pi)$MHz, while detuning is maintained at $\Delta=-2\Omega_0$ for the paramagnetic down-spin ordering ($|h_z^{C,V}| \gg J$) of Phase I. In the second stage ($0.1<t/t_f<0.9$), the detuning is swept from $-2\Omega_0$ to $2\Omega_0$, while the Rabi frequency is maintained at $\Omega_0$. In the final stage ($0.9<t/t_f<1$), the detuning is maintained at $\Delta_f=2\Omega_0$ (for anti-ferromagnetic ordering, $0<\Delta_f<3U$) and the Rabi frequency is adiabatically turned off from $\Omega_0$ to $\Omega_f=0$. The total operation time $t_f=(2\pi)3.2/\Omega_0=2.9$~$\mu$s is chosen long enough to identify the max-population states and short compared to the coherence time of 10~$\mu$s~\cite{Kim2020}. After the time evolutoin, the resulting spin configuration is detected and the procedure is repeated until the probability distribution of all spin configurations is obtained. 
The final Hamiltonians $\hat H_G(U,\Delta_f=2\Omega_0)$ are chosen at various phases in Fig.~\ref{fig2}: {\textcircled{\small 1}} and {\textcircled{\small 2}} for Phase III of $G_{10}$ and $G_{22}$, respectively; {\textcircled{\small 3}} for Phase IV of $G_{14}$; and {\textcircled{\small 4}} and {\textcircled{\small 5}} for Phase V of $G_{14}$.
\begin{figure*}[t]
\centerline{\includegraphics[width=1.00\textwidth]{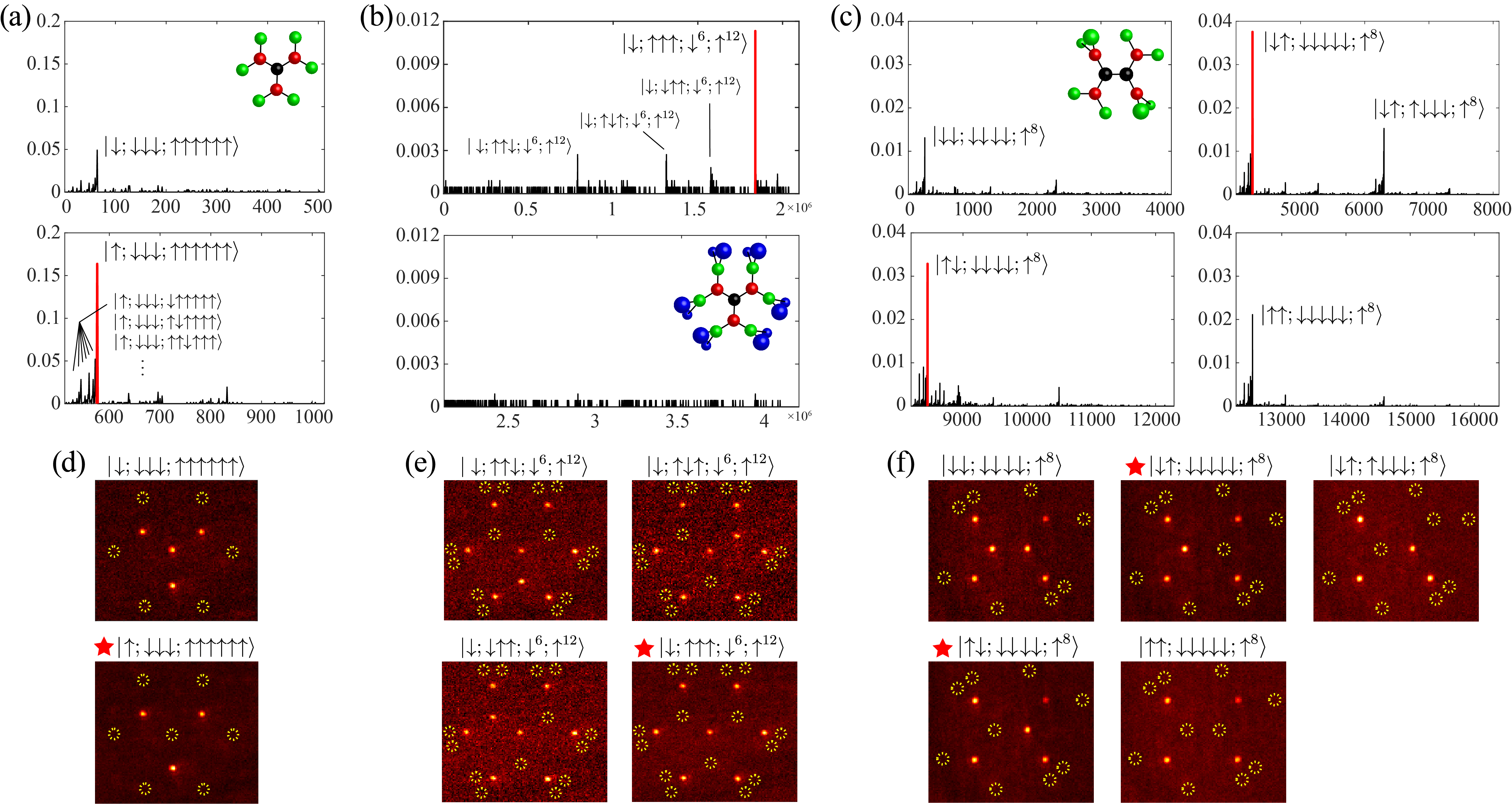}}
\caption{%{\bf Quantum annealing of Ising-spin Cayley trees.} 
Probability distribution of all spin configurations in enumerated bare-spin basis: (a) $G_{10}$ Cayley-tree atoms measured at {\textcircled{\small 1}} in Phase III and accumulated over 672 events. (b) $G_{22}$ at  {\textcircled{\small 2}} in Phase III (2208 events). (c) $G_{14}$ at  {\textcircled{\small 4}} in Phase V (5113 events). Max populations are ground states (highlighted in red) and smaller peaks are identified near-ground excited states. (d,e,f) Fluorescence images of ground-state ($\ket{\downarrow}$) atoms in characteristic spin configurations.}
\label{fig3}
\end{figure*}
\noindent

Measured probability distributions are shown in 
Figs.~\ref{fig3}(a-c), which plot the probability distributions, respectively measured at {\textcircled{\small 1}}, {\textcircled{\small 2}}, and {\textcircled{\small 4}}, for all spin configurations of the Cayley-tree Ising spins. Spin configurations are represented in the bare-atom basis with binary enumeration of  $\ket{\downarrow}=\ket{0}$ and $\ket{\uparrow}=\ket{1}$~\cite{Kim2018}. After the quantum annealing, the wavefunction of the atoms is expectedly driven near the ground state of $\hat H_G(U, \Delta_f)$, so the corresponding spin configuration is measured with a high probability. In Fig.~\ref{fig3}(a), Phase III of $G_{10}$ is probed at {\textcircled{\small 1}}. The observed max-population state is $\ket{\uparrow;\downarrow\downarrow\downarrow;\uparrow\uparrow\uparrow\uparrow\uparrow\uparrow}=\ket{2^{9}+2^5+2^4+2^3+2^2+2^1+2^0}=\ket{575}$, which is the anti-ferromagnetic phase of shell-by-shell alternating spins, agreeing with the expected ground state of $\hat H_{G_{10}}$ in Phase III. Likewise, in Figs.~\ref{fig3}(b), Phase III of $G_{22}$ is probed at {\textcircled{\small 2}} and the observed $\ket{\downarrow;\uparrow^3;\downarrow^6;\uparrow^{12}}=\ket{1839103}$ agrees with the ground-state of $\hat H_{G_{22}}$ in Phase III. In Fig.~\ref{fig3}(c), which shows the measurement of $G_{14}$ at  {\textcircled{\small 4}} in Phase V, high populations are observed for $\ket{\downarrow\uparrow;\downarrow^4;\uparrow^8}=\ket{4351}$ and $\ket{\uparrow\downarrow;\downarrow^4;\uparrow^8}=\ket{8447}$, agreeing with the ground superposition state (of center spins in $(\ket{\uparrow\downarrow}+\ket{\downarrow\uparrow})/\sqrt{2}$) in Phase V of $\hat H_{G_{14}}$. Atom images (the fluorescence of ground-state atoms) of characteristic spin configurations are shown in Figs.~\ref{fig3}(d,e,f), including the max-population states (of star marks). 

\begin{figure}[tb]
\centerline{\includegraphics[width=0.50\textwidth]{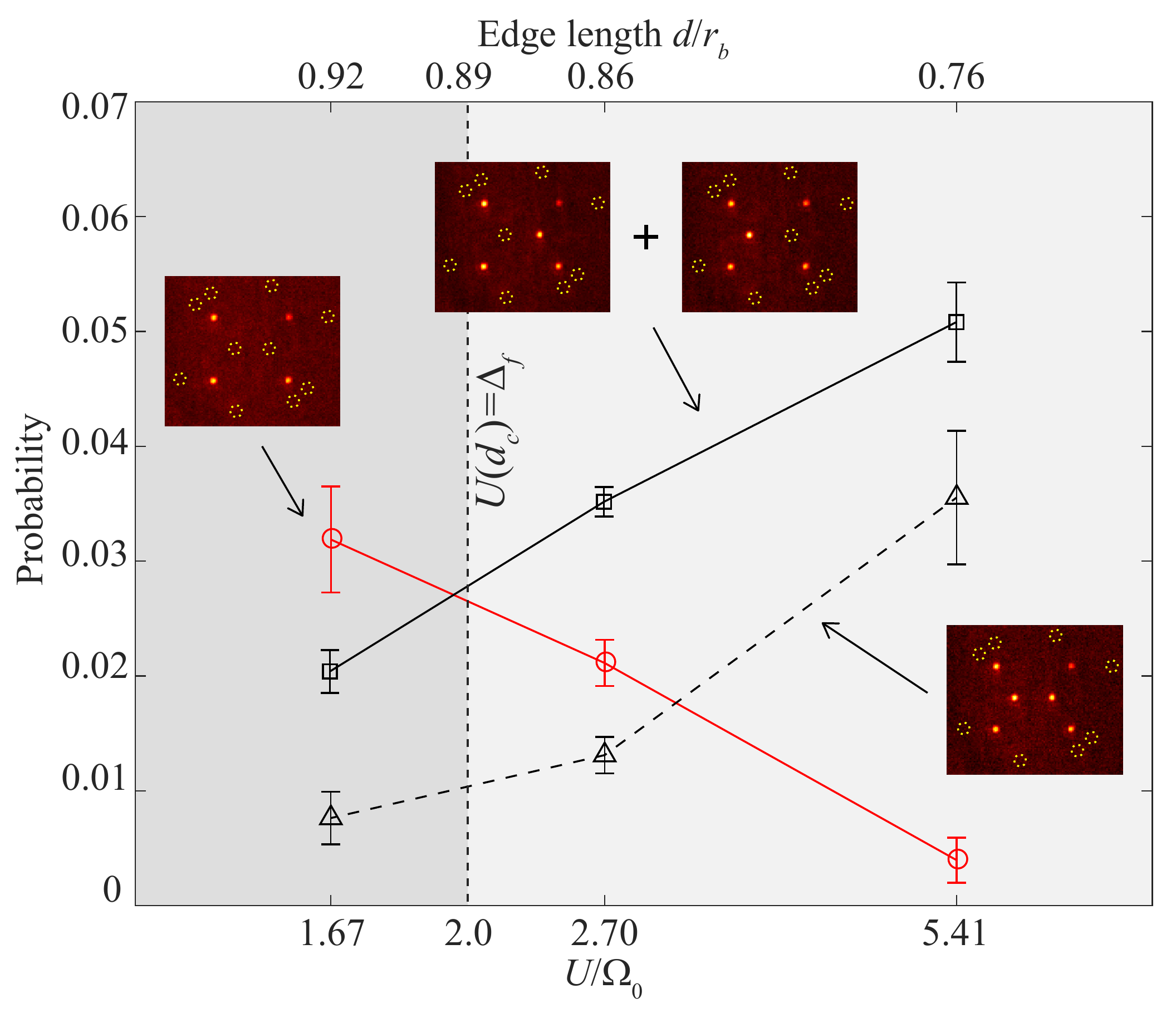}}
\caption{Probabilities of the low-energy states of Cayley-tree Hamiltonian, $\hat H_{G_{14}}(U(d), \Delta_f=2\Omega_0)$, measured for three different edge lengths, $d/r_b=0.92$, 0.86, and 0.76, which correspond to  {\textcircled{\small 3}}, {\textcircled{\small 4}}, and {\textcircled{\small 5}}, respectiveily, in the $G_{14}$ phase diagram in Fig.~\ref{fig2}(b).}
\label{fig4}
\end{figure}

Phase IV of $G_{14}$ is also probed acrossing the IV-V phase boundary, with three different Cayley trees of respective edge lengths $d/r_b=0.76$, 0.86, and 0.92, which correspond to {\textcircled{\small 5}}, {\textcircled{\small 4}}, and {\textcircled{\small 3}} in the phase diagram. In Fig.~\ref{fig4}, the probabilities of three high-population states, of respective center-spin configurations $\ket{\uparrow\uparrow}$, $\ket{\downarrow\downarrow}$, and $(\ket{\uparrow\downarrow}+\ket{\downarrow\uparrow})/\sqrt{2}$, are plotted. The max-populated state changes from $\ket{\uparrow\uparrow}$ ($d<d_c=0.89r_b$, Phase IV) to $(\ket{\uparrow\downarrow}+\ket{\downarrow\uparrow})/\sqrt{2}$ ($d>d_c$, Phase V), agreeing with the phase boundary given by $U(d_c)=\Delta_f$. It is noted that the higher-order long-range couplings which are ignored in Eq.~\eqref{HG} play little role in the tested parameter region and that other ground states, of center spins $\ket{\downarrow\uparrow}$, $\ket{\uparrow\downarrow}$, and $(\ket{\uparrow\downarrow}-\ket{\downarrow\uparrow})/\sqrt{2}$) are forbidden by the Hamiltonian symmetry. Also, the first excited state changes from $(\ket{\uparrow\downarrow}+\ket{\downarrow\uparrow})/\sqrt{2}$ to $\ket{\downarrow\downarrow}$ and then to $\ket{\uparrow\uparrow}$ from left to right, in accordance with numerical calculation.

\section{N\'{e}el's order formation dynamics}
\begin{figure}[t]
\centerline{\includegraphics[width=0.49\textwidth]{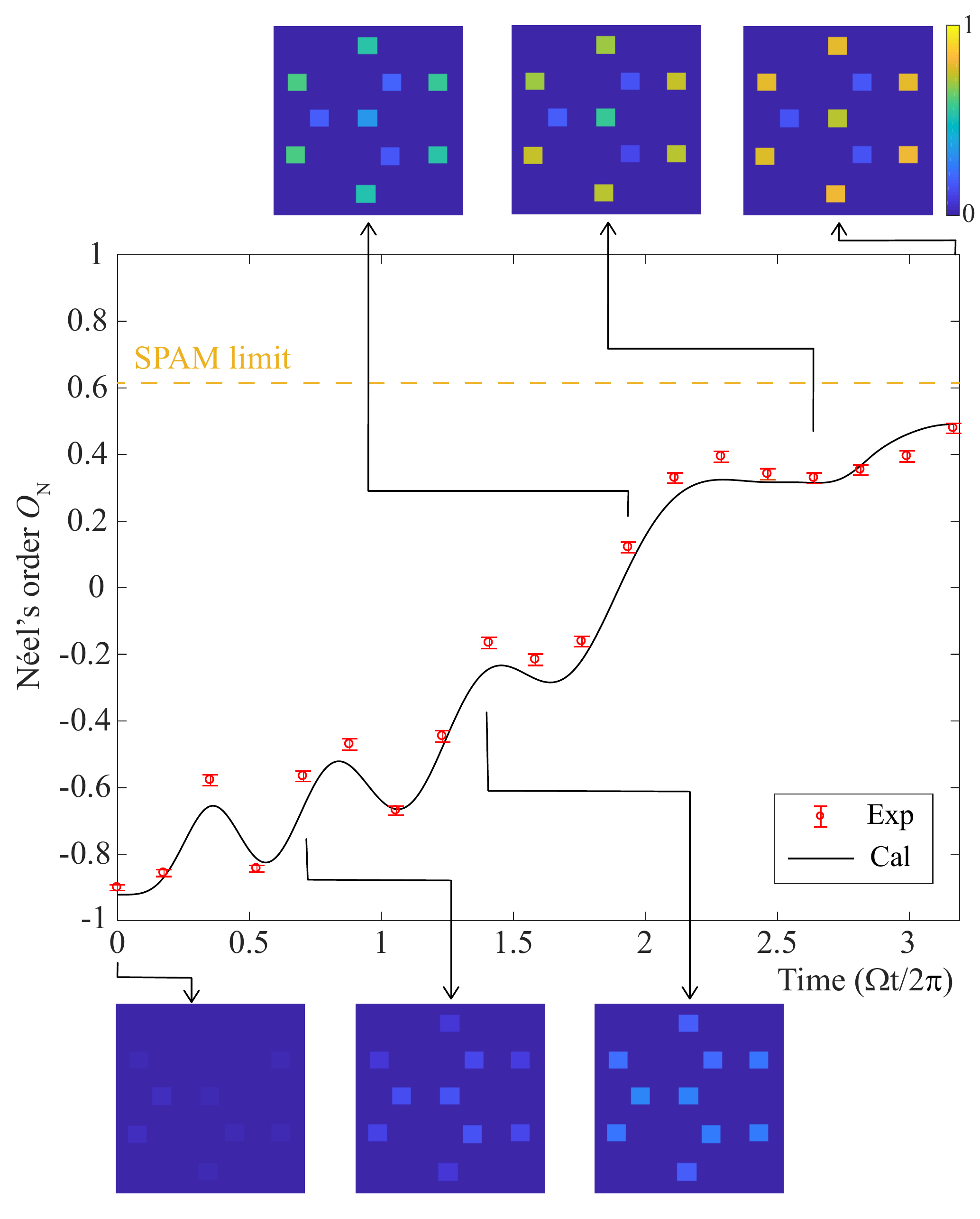}}
\caption{%{\bf Adiabatic N\'{e}el's order formation}. 
Quantum annealing dynamics of the N\'{e}el's order in the $G_{10}$ Cayley tree are probed as a function of the evolution time and compared with numerical calculations. At chosen times, the Rydberg-state probabilities, $\langle{\hat n}\rangle_j$, of all atoms ($j =1, \cdots, 10$) are shown, being plotted at their respective atom sites.}
\label{fig5}
\end{figure}

\noindent
Furthermore, we measure the phase formation dynamics during the quantum annealing process. Figure~\ref{fig5} shows the time evolution of N\'{e}el's order, defined by $O_N\equiv -\sum_{(i,j)\in E} \langle\hat\sigma^{(i)}(t) \hat\sigma^{(j)}(t)\rangle /||E||$, where $||E||$ is the number of edges, along with the up-spin probabilities of individual atoms in $G_{10}$. With the snap-shot measurements, the adiabatic order formation is clearly observed from the initial paramagnetic phase to the final anti-ferromagnetic phase. The oscillatory behavior is attributed to the finite size. In comparison, a numerical calculation (solid lines) is performed with Lindblad master equations, taking into account state-preparation-and-detection (SPAM) errors ($P(\ket\downarrow|\ket\uparrow)=0.18$, $P(\ket\uparrow|\ket\downarrow)=0.02$), individual dephasing ($\sim$36~kHz $\ll \Omega_0$) due to the spontaneous decay during Rydberg excitations, and collective dephasing ($\sim$3~kHz) from laser phase noise~\cite{Lee2019}. The numerical calculation in accordance with the observed maximal N\'{e}el's order of $O_N (t_f) = 0.48(2)$ indicates that errors are largely due to the SPAM errors accumulated for $N=10$ atoms. The calculation also suggests that the maximal N\'{e}el's order before measurements was $O_N(t_f)=0.82$ and that the ground-state probability was 61\% through the quantum annealing. 

\section{Experimental details} \label{technical}
\noindent
The above experiments were performed with a Rydberg-atom quantum simulator previously reported elsewhere~\cite{Kim2018,Lee2019,Jo2020,Kim2020}. In the quantum simulator, rubidium ($^{87}$Rb) atoms were initially prepared in the hyperfine ground state, $\ket{\downarrow}=\ket{5S_{1/2},F=2, m_F=2}$, and trapped with optical tweezers. A spatial light modulator (SLM, Meadowlark ODPDM512) was used to create the 3D array of $2N$ optical tweezers and an electrically focus-tunable lens (EL-16-40-TC from Optotune) verified the positions of captured atoms in each atom plane. The SLM was computer-programmed with weighted Gerchberg-Saxton (w-GS) algorithm~\cite{DiLeonardo2007,Kim2019} so that the resulting electric field near each target site $(x, y, z)$ was created as $E(x, y, z) = E_0 \sum_{X,Y}e^{i\Phi(X,Y)}e^{-iT}$, where $X,Y$ are the SLM coordinates on the Fourier plane of the optical tweezers, $\Phi(X,Y)$ is the SLM phase pattern, and $T= {2\pi (x X+y Y)}/{f\lambda } + {\pi z(X^2+Y^2)}/{ f^2 \lambda}$ is the transfer kernel. The phase pattern $\Phi$ (for $2N$ optical tweezers) was obtained with the w-GS condition, given by $\Phi(X,Y)={\rm arg}(\sum_{j=1}^{2N} {w_j E_j(x,y,z)}e^{iT_j}/ {\abs{E_j(x,y,z)}})$, where each weighting factor $w_j$ for $j^{\rm th}$ optical tweezer was optimized through adaptive iterations. The number of the w-GS iterations was about 5, taking about 20~ms time to generate an array of 50 tweezers. Determinstic rearrangement of $N$ atoms to target sites, about 20~$\mu$m apart from the initial atom reservoir, was performed with consecutive 45 frames of moving traps programmed with phase induction~\cite{Kim2019}. Over 90\% target-occupation probabilities were achieved in 900-ms reconfiguration time and all measurements were performed with defect-free arrangements.

After the atom array was prepared, the optical tweezers were temporarily turned off and the quantum annealing was proceeded. The atoms were excited to the Rydberg state, $\ket{\uparrow}=\ket{71S_{1/2},m_J=1/2}$, via the off-resonant intermediate state, $\ket{i}=\ket{5P_{3/2}, F'=3, m_F'=3}$. Two lasers (780-nm and 480-nm lasers for $\ket{\downarrow}\rightarrow\ket{i}$ and $\ket{i}\rightarrow\ket{\uparrow}$) were used for the two-photon transition~\cite{Kim2020}.  $\Omega(t)$ and $\Delta(t)$ in $H(t)$ were programmed with a RF synthesizer (Moglabs XRF, 10~MHz) of frequency (780 nm) and amplitude (480 nm) modulations. The modulation ranges were $0\le\Omega(t)\le\Omega_0$ and $-2\Omega_0\le\Delta(t)\le2\Omega_0$, where the max Rabi frequency was $\Omega_0=\Omega_{780}\Omega_{480}/(2\Delta')=1.1$~$(2\pi)$MHz given by $\Omega_{780}=75$~$(2\pi)$MHz, $\Omega_{480}=19$~$(2\pi)$MHz, and $\Delta'=660$~$(2\pi)$MHz (the intermediate detuning). The van der Waals coefficient~\cite{Weber2017} was $C_6=(2\pi)1004$~GHz$\times{\mu m^6}$ for $\ket{71S}$ Rydberg-state atoms and the Rydberg blockade radius was $r_b=({C_6/\Omega_0})^{1/6}=9.8~{\rm \mu m}$. 
The frequency error due to AC Stark shift was small, below 140 kHz, mainly caused by the 480-nm laser.  After the quantum annealing, the optical tweezers were turned back on and the atom states in the ground state ($\ket \downarrow $) were measured whether they survived ($\ket \downarrow $) or not ($\ket \uparrow$). Each measurement in Fig.~\ref{fig3} was repeated by 672, 2208, and 5113 times respectively for $G_{10}$, $G_{22}$, and $G_{14}$ to obtain the accumulated probability distributions.

\section{Conclusion} 
\noindent
We have explored the possibilites of Rydberg-atom quantum simulators towards high-dimensional qubit connection programming. With up to $N=22$ rubidium single atoms arranged in three-dimensional space, we have programmed Ising Hamiltonians on three different Cayley tree graphs. The anti-ferromagnetic phase in regular Cayley trees and frustrated competing ground-states in a dual-center Cayley tree are directly observed, showing good agreement with model calculations. There are many hopes to utilize the potentials of Rydberg-atom quantum simulators in large-scale implementations and high-dimensional configurability~\cite{LHZ2015, DiehlNP2011, Glaetzle2017, Pichler2018, Serett2020,Qiu2020}. It is hoped that 3D-qubit configurations of Rydberg-atom quantum annealers
shall be useful for efficient and programmable quantum optimization problems.

\begin{acknowledgements}
This research was supported by Samsung Science and Technology Foundation (SSTF-BA1301-52), National Research Foundation of Korea (NRF) (2017R1E1A1A01074307), and Institute for Information \& Communications Technology Promotion (IITP-2018-2018-0-01402). 
\end{acknowledgements}

\end{document}